\documentclass[aps,prl,twocolumn,groupedaddress,showpacs]{revtex4}

\usepackage{epsfig}
\usepackage{amssymb}

\begin{document}

\title{Atom laser divergence}
\author{Y.\ Le~Coq}
\author{J.\ H.\ Thywissen}
\author{S.\ A.\ Rangwala}
\author{F.\ Gerbier}
\author{S.\ Richard}
\author{G.\ Delannoy}
\author{P.\ Bouyer}
\author{A.\ Aspect}
\affiliation{Laboratoire Charles-Fabry de l'Institut d'Optique, 
UMRA 8501 du CNRS, 91403 Orsay, France}
\homepage{http://atomoptic.iota.u-psud.fr}
\date{\today}

\begin{abstract}
%
%
We measure the angular divergence of a quasi-continuous,
rf-outcoupled, free-falling atom laser as a function of the
outcoupling frequency.
The data is compared to a Gaussian-beam model of laser propagation
that generalizes the standard formalism of photonic lasers. Our
treatment includes diffraction, magnetic lensing, and interaction
between the atom laser and the condensate.
We find that the dominant source of divergence is the
condensate-laser interaction.
\end{abstract}
\maketitle
%
%

%
A dilute gas of atoms condensed in a single quantum state of a
magnetic trap \cite{cornell,ketterle,hulet,kleppner,hebec,gasou} is the
matter-wave analog of photons stored in an optical cavity.  A further
atomic parallel to photonics is the ``atom laser''
\cite{mewes,anderson,hagley,bloch}, a coherent extraction of atoms
from a Bose-Einstein condensate.
Atom lasers are of basic interest as a probe of
condensate properties \cite{2laser} and as a coherent source of atoms
\cite{fringe}. Furthermore, just as optical lasers greatly exceeded
previous light sources in brightness, atom lasers will be useful in
experiments that simultaneously require monochromaticity, collimation,
and intensity, such as holographic atom lithography \cite{shimizu},
continuous atomic clocks \cite{clocks}, and coupling into atomic
waveguides \cite{wgload}.

\begin{figure}[b]
\epsfig{file=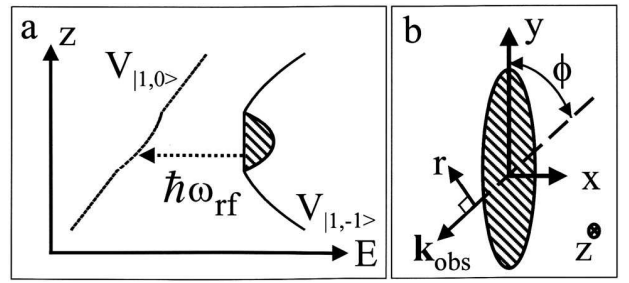, width=3.2in}
\caption{{\bf(a)} Schematic representation of outcoupling 
	from the trap potential $V_{\left|1,-1\right\rangle}$ to the
	untrapped $V_{\left|1,0\right\rangle}$ potential via an rf
	photon at $\omega_{\rm rf}$. The vertical direction is
	${\mathbf z}$. The curves shown include magnetic,
	gravitational, and mean field potentials. {\bf(b)} The
	wave-vector ${\bf k}_{\rm obs}$ for the absorptive imaging light
	beam at an angle $\phi$ with respect to the weak axis ${\bf
	y}$ of the condensate. The horizontal coordinate in the
	imaging plane is $r$. In both images, the condensate is
	cross-hatched.}  \label{fig:setup}
\end{figure}

%
Initial work on atom lasers has demonstrated several types of coherent,
pulsed output couplers \cite{mewes,anderson,hagley} as well as a
coherent narrow-band coupler \cite{bloch,narrowband}, which is the
type employed in our work. 
In this Letter, we address the nature of propagation of an atom laser
outcoupled from a condensate. Our experimental data show that the
laser beam is well characterized by a divergence angle. We measure
this angle versus radio-frequency (rf) outcoupler frequency, which
chooses the vertical extraction point of the atom laser from the
condensate (see Fig.\ \ref{fig:setup}a). In choosing the extraction
point, one chooses the thickness of the condensate to be crossed by
the extracted atoms, as well as the width of the atom laser beam at
the extraction plane.
To interpret this data, we use a formalism that generalizes the ABCD
matrices treatment of photonic lasers \cite{kogelnik,borde}. 
This treatment allows us to
calculate analytically the divergence of the laser due to diffraction,
magnetic lensing, and interactions with the condensate.
We find that, for our typical experimental conditions, the divergence
of the laser is primarily due to interaction between the atoms in the
laser and the atoms remaining in the condensate. We describe this
interaction as a thin lens.  Note that the existence of such an
interactive lensing effect is in stark contrast to a photonic laser,
since photons do not interact. Interactions were estimated to be
similarly important for pulsed atom lasers \cite{mewes,hagley}. In our
case, the divergence is also magnified by a thick-lens-like potential
due to the quadratic Zeeman effect of the magnetic trapping fields.
%

%
The technique we use to obtain Bose-Einstein condensates with
$^{87}$Rb is described in detail elsewhere \cite{orsayBEC}. Briefly, a
Zeeman-slowed atomic beam loads a MOT in a glass cell. Typically
10$^8$ atoms are transferred to a magnetic trap, which is subsequently
compressed to oscillation frequencies of $\omega_x = \omega_z = 2\pi
\times 144$\,Hz and $\omega_y = 2\pi \times 9$\,Hz$=\lambda \omega_z$ in 
the quadrupole and dipole directions, respectively, where $\mathbf{z}$
is vertical. The Ioffe-Pritchard trap is created with an iron-core
electromagnet, with a typical quadrupole gradient of
11.7\,T$\cdot$m$^{-1}$, and an uncompensated dipole bias field of
$B_0=$5.4\,mT. A 40\,s, rf-induced evaporative cooling ramp in the
compressed trap results in condensates of typically $5 \times 10^5$
atoms, with a 15\% rms shot-to-shot variation.

\begin{figure}
\epsfig{file=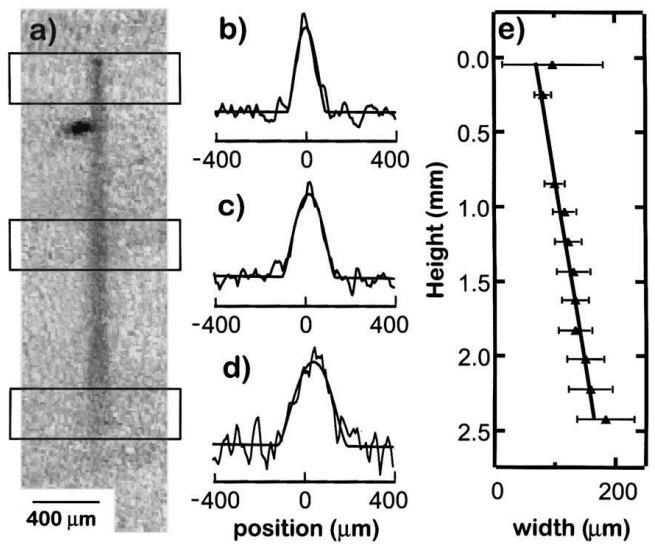, width=3.4in}
\caption{Typical continuous atom laser output. {\bf(a)} Absorptive 
	image of laser, after 10\,ms of output coupling at 38.557\,MHz
	and 6\,ms of free flight. The condensate (the darkest area of
	image) is displaced from the beginning of the laser because of
	a magnetic kick separating atoms in the $m_F = -1$ and $m_F =
	0$ states. {\bf(b)}--{\bf(d)} Examples
	of absorption profiles of the atom laser, taken from the three
	regions boxed in {(a)}. Each profile is fit to find a width
	$R_n$ and density $\rho_n$. {\bf(e)} The divergence angle of a
	single laser is found with a linear fit to the series of
	measured widths $\{R_n\}$.}
\label{fig:laser}
\end{figure}

%
Atom lasers are created by applying a rf field at about 38.6\,MHz, to
transfer condensate atoms from the trapped $|F,m_F\rangle =
|1,-1\rangle$ state to the weakly anti-trapped $|1,0\rangle$ state,
which also falls under gravity (see Fig.\
\ref{fig:setup}a). The rf field is weak (approximately 0.1\,$\mu$T)
and applied for a relatively long duration ($t_{\rm{oc}}=10$\,ms)
\cite{narrowband}.
There is no significant coupling to the
$|1,+1\rangle$ state because the $F=1$ sublevel transitions are split
by 0.8\,MHz due to the nonlinear part of the Zeeman effect at 5.4\,mT.
To measure the spatial distribution of the atom laser, we take an
absorptive image with a pulse of resonant light 6\,ms after the moment
when the rf outcoupling ends and the trap is turned off. As depicted
in Fig.\ \ref{fig:setup}b, the image is taken at a $\phi=55^\circ$
angle from the weakly confining $\mathbf{y}$ axis of the trap. Figure
\ref{fig:laser}a shows a typical image of an atom laser.

The gravitational sag $g/\omega_z^2$ shifts the entire condensate from
the center of the magnetic trap to a region where iso-field surfaces
are approximately planes of constant height $z_0$ across the
condensate \cite{bloch}.
The relation between rf frequency $\omega_0 + \delta_{\rm{rf}}$ 
and coupling height $z_0$ is
\begin{equation}
\delta_{\rm rf} = - \Delta \frac{z_0}{R_z} 
- \frac{\mu}{\hbar}\left(1 - \frac{z_0^2}{R_z^2}\right)
\label{eq:delta}
\end{equation}
where $\Delta = M g R_z/\hbar$ is the spectral half-width of the
condensate, $M$ is the mass of the atom, $g$ is gravitational
acceleration, and $R_z$ is the Thomas-Fermi \mbox{(TF)}
radius \cite{dalfovo} of the condensate along $\mathbf{z}$. The radius
$R_z$ is $\sqrt{2
\mu/M \omega_z^2}$, and $\mu$ is the chemical potential $(\hbar
\bar{\omega}/2) (15 a_{11} N)^{2/5} (M
\bar{\omega}/\hbar)^{-1/5}$, where $N$ is the number of atoms in the
condensate, $a_{11} = 5.67$\,nm is the s-wave scattering length
between $^{87}$Rb atoms in the $|1,-1\rangle$ state \cite{hall}, and
$\bar{\omega}^3=\omega_x \omega_y \omega_z $. In Eq.\
(\ref{eq:delta}), we have chosen $\omega_0$ such that $\delta_{\rm rf}
= \pm \Delta$ when $z_0 = \mp R_z$.
For our experimental parameters, $\mu/\hbar \ll \Delta$, so
$\delta_{\rm rf}$ is roughly linearly dependent on $z_0$, with slope
$-Mg/\hbar$. Therefore, in choosing the coupling frequency, we choose
the height within the condensate at which the laser is sourced.

%
Short-term (ms-scale) stability of the bias field is verified by the
continuity of flux along the laser (Fig.\ \ref{fig:laser}a). We find
that the shot-to-shot stability and reproducibility of the bias field,
between each 80\,s cycle of trapping, cooling, and condensation, is
about $\pm 0.4$\,$\mu$T or $\pm 3$\,kHz. This stability is sufficient
to scan through the condensate spectral width of about 20\,kHz. In one
out of five runs (a run is a set of about 10 cycles), the data were not
self-consistent, which we attribute to a larger ($>$1\,$\mu$T) bias
field fluctuation during that run. We could maintain this bias field
stability, typically one part in $10^4$ \cite{yoke}, by using either a
low-noise power supply or a battery, since only 1.3\,A is used to
energize the dipole coils during evaporation and outcoupling.
%

%
We analyze the images (such as Fig.\ \ref{fig:laser}a) to measure the
flux and divergence of the outcoupled laser. Figures \ref{fig:laser}b
through \ref{fig:laser}d
shows the first step in the analysis, a series of fits to the
transverse spatial profile of the laser, as measured at several
heights $z_n$, averaging across $\pm140$\,$\mu$m.  The column density
profile is fit at each $z_n$ by $\rho_n(1-r^2/R_n^2)^{3/2}$, where $r$
is the horizontal coordinate in the image plane (see Fig.\
\ref{fig:setup}b), $\rho_n$ is the peak column density, and $R_n$ is
the width. This fit function would be the rigorously correct function
for an atom laser without divergence
\cite{gerbier} or for a free condensate undergoing mean field
expansion \cite{profile}; here, we observe empirically that the fits
are good. From the integral of the column densities at each height, we
determine the output flux $F$ with a fit to $F/\sqrt{2g(z-z_0)}$, a
form that assumes constant flux, purely gravitational acceleration,
and a density that decreases with the inverse of the classical
velocity, valid when $\left|z - z_0 \right| \gg \ell_g$, where $\ell_g
= [\hbar^2/(2 g M^2)]^{1/3}$ is the gravitational length. Finally,
from the series $\{R_n\}$ and $\{z_n\}$, we determine a geometric
expansion angle by a linear fit (see Fig.\ \ref{fig:laser}e).  We will
discuss below why one would expect a linear rather than parabolic
shape for a laser falling under gravity.

\begin{figure}
\epsfig{file=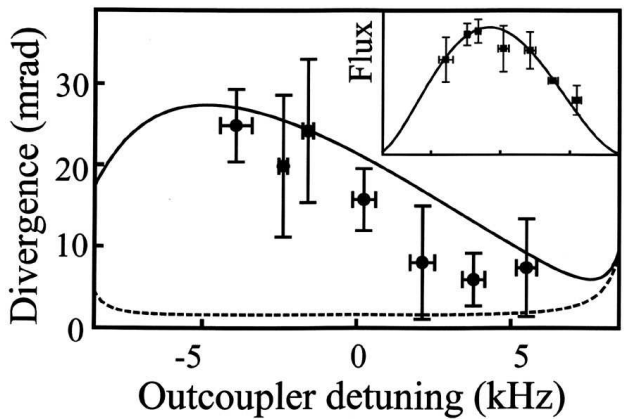, width=3.25in}
\caption{Divergence (half-angle) versus output coupler detuning
        from the condensate center. Experimental points are compared
        with the theoretical calculation (solid line). The dashed line
        represents the same calculation, but excluding the effects of
        the condensate-laser interaction.  {\bf Inset:} Output flux
        (in arbitrary units) versus detuning, with the same frequency
        scale as the main figure.}
\label{fig:angle}
\end{figure}

%
We repeated the above imaging, and analysis for atom lasers coupled at
a variety of rf coupling frequencies.
Figure \ref{fig:angle} shows the averaged half-angle divergence versus
outcoupler detuning, from 15 laser images.  We see that the divergence
clearly decreases at higher $\delta_{\rm{rf}}$, corresponding to lower
initial heights $z_0$. These divergence data will be analyzed in more
detail below.
The inset of Fig. \ref{fig:angle} shows laser
flux $F$ versus $\delta_{\rm{rf}}$, and is compared to
\begin{equation}
\frac{F}{F_0} = \left(1 - \frac{z_0^2}{R_z^2}\right)^2
\left[1 - \frac{2 \mu}{\hbar \Delta} \frac{z_0}{R_z} + 
\frac{2 \mu^2}{3 \hbar^2 \Delta^2} (1 - \frac{z_0^2}{R_z^2}) \right],
\label{eq:flux}
\end{equation}
where $F_0$ is the peak flux, and Eq.\ (\ref{eq:delta}) defines the
relation between $z_0$ and $\delta_{\rm rf}$. Equation (\ref{eq:flux})
assumes the laser flux is simply proportional to the linear density of
the condensate at the coupling point.
For the solid curve shown in the inset of Fig.\ \ref{fig:angle}, we
have used $N=4 \times 10^5$ to give $\Delta = 2 \pi \times 9.1$\,kHz
and $R_z = 4.2$\,$\mu$m.
The peak flux is measured to be about $1 \times 10^{7}$\,s$^{-1}$ by
the reduction in condensate number, as in \cite{bloch}. This is in
agreement with the theory of \cite{gerbier}, given the applied field
strength and uncertainty in the atom number and field polarization
\cite{narrowband}.
%

%
There are several possible sources of divergence of the atom laser,
including diffraction, magnetic lensing, and interactions both within
the laser and between the laser and the condensate. In order to
understand the divergence with a simple analytical model, we make
several approximations: 1) that the interactions between atoms {\em
within} the laser are not significant, valid in the low-flux limit; 2)
that the roughly parabolic density profile of the atom laser can be
approximated by a Gaussian; and 3) that we can use stationary
solutions of the Schr\"{o}dinger Equation with a paraxial-type of
approximation, in which the fast degrees of freedom \cite{gerbier} are
decoupled from the slow evolution of the transverse degrees of
freedom, as in \cite{wallis}.
We follow a Gaussian optics treatment similar to that of photonic
lasers \cite{kogelnik}: the spatial distribution in $\mathbf{x}$ and
$\mathbf{y}$ described by the wavefunction $\Psi(x,y,t) = \Psi_0
\exp{\left[-i P(t) + i x^2/2 q_x(t) + i y^2/2 q_y(t)\right]}$, where
$P(t)$ describes the overall phase and amplitude, and beam parameters
$1/q_{x,y}(t) = i/w^2_{x,y}(t) + c_{x,y}(t)$ describe the widths
$w_x$ and $w_y$ and curvatures $c_x$ and $c_y$ of the beam. The
observable width is $w(t) = (w^2_x(t) \cos^2{(\phi)} + w^2_y
\sin^2{(\phi)})^{1/2}$, where $\phi$ is the observation angle 
(Fig.\ \ref{fig:setup}b). The initial widths $w_x^2(0) = 2(R_z^2 -
z_0^2)/5$ and $w_y(0) = w_x(0)/\lambda$ give the same rms width as
would an initial Thomas-Fermi density profile.

The beam parameters $q_x$ and $q_y$ follow an ``ABCD law'' similar to
that for an photon laser beam: $q' = (A q + B)/ (C
q + D)$, where the coefficients $A$, $B$, $C$, and $D$ are the four
elements in a matrix which transform ray vectors $\langle x, p/\hbar
\rangle$ to $\langle x', p'/\hbar \rangle$ according to classical
equations of motion in the same potential.
In the following paragraphs, we will calculate the ABCD matrices for
interaction with the condensate, propagation with the magnetic trap
on, and free flight after the trap is turned off.  Note that even
though the ray matrices can be derived using classical equations of
motion, their application in the Gaussian beam formalism includes
diffraction.

The mean-field interaction potential between an atom in the laser and
the condensate is $U_I({\mathbf r}) = g_{01} \rho_{c}({\mathbf r})$, where
$g_{01}$ is the s-wave coupling strength $4 \pi \hbar^2 a_{01}/M$
between atoms in the $|1,0\rangle$ state and the trapped
$|1,-1\rangle$ state, and $\rho_{c}({\mathbf r})$ is the condensate
density. Here we use $a_{01} = a_{11}$. We calculate the action of
this potential treating it as a lens, and using the thin lens
approximation that each trajectory is at its initial transverse
position $\left\langle x_0, y_0 \right\rangle$.
In the Thomas-Fermi limit, the potential $U_I({\mathbf r})$ is
quadratic, and thus gives an impulse $\left\langle \Delta p_x, \Delta
p_y \right\rangle  = \left(2 \mu\, t_1(x_0,y_0)/R_z^2\right) \left\langle x_0,
\lambda^2 y_0 \right\rangle$ after an interaction time $t_1$.
The on-axis power of the lens is given for $x_0 \ll R_x$ and $y_0 \ll
R_y$, with which assumption we find $t_1^2 \approx 2(R_z+z_0)/g$. This
gives the thin lens ABCD matrix for the $\mathbf x$ direction
\begin{equation}
M_{1x}(z_0) =
\left( \begin{array}{cc} 1 & 0 \\
\frac{2 \mu}{\hbar R_z^2}\sqrt{2 (R_z + z_0)} & 1 \end{array} \right).
\label{eq:ABCD1}
\end{equation}
When applied to the beam parameter $q_x$, the nontrivial term in Eq.\
(\ref{eq:ABCD1}) is the wavefront curvature added to the beam. A
similar ray matrix $M_{1y}(z_0)$ transforms $q_y$, but with a
curvature term multiplied by $\lambda^2$.
For both $M_{1x}(z_0)$ and $M_{1y}(z_0)$, the curvature is positive
for all $z_0$, since the interaction is always repulsive. A positive
curvature corresponds to an expanding wave, and thus the condensate
with repulsive interactions ($g_{01}>0$) is always a {\it diverging}
lens.
%

%
When the atom laser falls \cite{qzesplit}, it evolves in the
anti-trapping potential due to the quadratic Zeeman effect of the
$|1,0\rangle$ state, $U_{\rm QZE}({\mathbf r}) = -\mu_B B^2({\mathbf
r})/B_{\rm HF}$, where $\mu_B$ is the Bohr magneton, and $B_{\rm HF} =
0.4883$\,T is the hyperfine splitting in magnetic units. 
We can neglect $y$-dependent and higher-order terms, since they
are several orders of magnitude smaller for the fields in our
experiment, to get $U_{\rm QZE}({\mathbf r}) \approx -\mu_B
B_0^2/B_{\rm HF} - M \Omega^2 (x^2 + z^2)/2$, where $\Omega = 2 \pi
\times 30.3\pm0.1$\,Hz. Below, we will return to
evolution in ${\mathbf y}$.
The classical motion of a particle in an inverted quadratic potential
is given by hyperbolic functions. In the vertical direction, an 
elongation of the laser is evident:
we observe a length of 1.84$\pm$0.09\,mm, while with gravity alone, one would
expect a length of 1.33\,mm after 10\,ms of coupling and 6\,ms of free
flight. Including $U_{\rm QZE}({\mathbf r})$ and our measured trap parameters,
we calculate 1.87\,mm, in agreement with our observations. 
The ray matrix for the transverse $\mathbf{x}$ direction is
\begin{equation}
M_2(t_2) =
\left( \begin{array}{cc} \cosh(\Omega t_2) &
			\frac{\hbar}{M \Omega} \sinh(\Omega t_2) \\ 
			\frac{M \Omega}{\hbar} \sinh(\Omega t_2) &
			\cosh(\Omega t_2) \end{array} \right),
\label{eq:ABCD2}
\end{equation}
where $t_2$ is the time of evolution, ranging between $0$ and
$t_{\rm{oc}}$. This interaction is a {\em thick} lens, since $\Omega
\,t_{\rm{oc}} > 1$, and thus there is sufficient time for the laser to
change diameter and curvature during its interaction.
During the same time, the beam parameter $q_y$ transforms by the free
flight matrix
\begin{equation}
M_{\rm FF}(t_2) = 
\left( \begin{array}{cc} 1 & \hbar t_2/M \\ 0 & 1 \end{array} \right),
\label{eq:ABCD3}
\end{equation}
which is the $\Omega \rightarrow 0$ limit of $M_2(t_2)$.
Note that it is due to the acceleration in both the $\mathbf{z}$ and
$\mathbf{x}$ directions that make the atom lasers well characterized
by an asymptotic expansion angle (see Fig.\ \ref{fig:laser}e): if
there were no acceleration in $\mathbf{x}$, the laser would have
parabolic borders.

%
The third and final transformation of the laser is free-flight
expansion between turning the trap off and observing the laser,
described by $M_{\rm FF}(t_{\rm F})$, where $t_{\rm F}$ is the time of
flight.
To find the width of the laser at any position, we evolve the initial
beam parameter $q_x(0) = -i w_x(0)^2$ and $q_y(0) = -i w_y(0)^2$ with
the elements of the matrix product $M_{\rm FF}(t_{\rm F}) M_2(t_2)
M_{1x}(z_0)$ and $M_{\rm FF}(t_2 + t_{\rm F}) M_{1y}(z_0)$,
respectively.  The angle is the ratio of the difference between the
observed rms beam size at $t_2=0$ and $t_2=t_{\rm{oc}}$ to the length
of the laser.  Note that this theory has no adjustable parameters
since the atom number, output flux, trap frequencies, interaction
strength, and bias field have all been measured.

%
Figure \ref{fig:angle} shows the calculated angle $\theta$ in comparison with
the data. The primary feature of this curve is its monotonous decrease
with increasing $\delta_{\rm rf}$ throughout the range of the
data. This trend is due to the condensate lens, as is demonstrated by
comparison to $\theta$ without the transformation $M_1$ (dashed line
in Fig.\ \ref{fig:angle}). Simply put, a laser sourced at lower
heights (greater $\delta_{\rm rf}$) interacts for less time with the
condensate. The quadratic Zeeman potential $U_{\rm QZE}$ acts to
magnify the divergence of the condensate lens,
increasing the slope in this region by roughly a factor of four.
%
%
Out of the range of our data, there are two more salient features: (1)
The angle $\theta$ decreases at $\delta_{\rm rf} < -5$\,kHz. This is
due to a reduction in initial width of lasers sourced from the very
top of the condensate, since divergence angle is
proportional to beam radius for a constant focal length. (2) The angle
$\theta$ increases for $\delta_{\rm rf}>5$\,kHz due to
diffraction. A combination of diffraction and interactions imposes the
minimum divergence possible on the atom laser, in our case
approximately 6\,mrad.

%
In conclusion, we have measured the divergence of an atom laser. We
demonstrate that in our case, interactions are a
critical contributor to the observed divergence.
The strong parallel between atom and photon lasers beams, both fully
coherent, propagating waves, is emphasized by the success of a model
obtained by generalization of the standard treatment of optical laser
beams.
The understanding of atom laser propagation provided by our
measurements and model provides a basic tool for future experiments
with atom lasers.
%
%
\begin{acknowledgments}
The authors thank W.\ D.\ Phillips, C.\ I.\ Westbrook, Ch.\ J.\
Bord\'e, M.\ Kohl, and S. Seidelin for comments. J.\ T.\ acknowledges
support from a Chateaubriand Fellowship. This work was supported by
the Direction G\'en\'eral de l'Armement (grant 99.34.050) and the
European Union (grant HPRN-CT-2000-00125).
\end{acknowledgments}
%
%

%


\begin{thebibliography}{99}
\bibitem{narrowband}{In this work, the coupling width is
uncertainty-limited at approximately $2 \pi \times 100$\,Hz, with
a the coupling rate is approximately $2 \pi \times
6$\,Hz. These widths are small compared to the chemical potential $h
\times 1.6$\,kHz.
%
We call these atom lasers ``narrow-band'' because the energetic width
of the output mode is limited by the coupling rate and/or the rf pulse
length (as in \cite{bloch}), instead of the spatial size of the
condensate (as in \cite{mewes,anderson,hagley}).}
\bibitem{yoke}{We attribute this field stability to the
magnetic shielding of the condensate by the iron yoke of the
electromagnet.}
\bibitem{qzesplit}{The potential $U_{\rm QZE}$ also acts within the
condensate, but is neglected transversely in the thin lens
approximation, and longitudinally because $R_z \ll g/\Omega^2$.}
\end{thebibliography}
\end{document}